\documentclass[aps,onecolumn,showpacs,pra,floatfix,superscriptaddress,longbibliography]{revtex4-1}
\usepackage{amsmath}
\usepackage{amsthm}
\usepackage{amsfonts}
\usepackage{bbm}
\usepackage{color}
\usepackage{graphicx}
\usepackage{dcolumn}
\usepackage{array}
\usepackage{float}
\usepackage{supertabular}
\usepackage{longtable}
\usepackage{txfonts}
\usepackage{wasysym}
\usepackage{cases}
\usepackage[T1]{fontenc}
\usepackage[latin1]{inputenc}
\usepackage{amssymb}
\usepackage[usenames,dvipsnames]{xcolor}
\usepackage{amstext}
\usepackage{latexsym}
\usepackage[colorlinks=true,citecolor=Cerulean,linkcolor=RubineRed,urlcolor=Cerulean]{hyperref}
\usepackage{enumitem}
\usepackage{epsfig}
\usepackage{dsfont}
\usepackage{mathrsfs}
\usepackage{arydshln,leftidx,mathtools}
\usepackage{color}
\usepackage{amsmath}

\begin{document}
\title{On generalized forces in higher derivative Lagrangian theory.}
\author{M. Beau}
\affiliation{Dublin Institute for Advanced Studies, School of Theoretical Physics, 10 Burlington Road, Dublin 4, Ireland}
\affiliation{Department of Physics, University of Massachusetts, Boston, Massachusetts 02125, USA}

\newcommand{\beq}{\begin{equation}}
\newcommand{\eeq}{\end{equation}}
\newcommand{\beqa}{\begin{eqnarray}}
\newcommand{\eeqa}{\end{eqnarray}}
\newcommand{\intf}{\int_{-\infty}^\infty}
\newcommand{\into}{\int_0^\infty}

\begin{abstract}
In this article, we introduce higher derivative Lagrangians
of this form $\alpha_1 A_{\mu}(x)\dot{x}^\mu$, $\alpha_2 G_{\mu}(x)\ddot{x}^\mu$, $\alpha_3 B_{\mu}(x)\dddot{x}^\mu$, $\alpha_4 K_{\mu}(x)\ddddot{x}^\mu$, 
$\cdots$, that generalize the electromagnetic interaction to higher order derivatives. We show that odd order Lagrangians describe interactions analog to electromagnetism while even order Lagrangians are similar to gravitational interaction.  
From this analogy, we formulate the concept of the generalized induction principle assuming the coupling between the higher fields $U_{(n),\mu}(x),\ n\geq1$
and the higher currents $j^{(n)\mu}=\rho(x)d^nx^{\mu}/ds^n$, where $\rho(x)$ is the spatial density of mass ($n$ even) or of electric charge ($n$ odd). In short, this article is an invitation to reflect on a generalization of the concept of force and of inertia. We discuss the implications of these paradigms more in depth in the last section of the paper. 
\end{abstract}

\maketitle

\tableofcontents

\section{Introduction}


In 1850, Mikha\"{i}l Ostrogradsky introduced the idea of higher derivative Lagrangians, see \cite{MecaG0}. The concept generalizes Newtonian mechanics to the case of forces that can depends explicitly on velocity, acceleration, and higher order derivatives, such as jerk, snap, etc... 
Generalized mechanics with higher derivative Lagrangians has been extensively studied and is ubiquitous. Some examples include, classical mechanics \cite{Borneas1,Borneas2,Anderson1,Anderson2,Alder}, quantum mechanics \cite{Hayes,Borneas3,Kleinert1}, relativistic mechanics \cite{Pisarski,Nesterenko}, the well-known Pais-Uhlenbeck oscillator \cite{Pais,Bolonek} and generalized electrodynamics \cite{PodolskyI,PodolskyII}, higher-derivative scalar field model \cite{Anisimov,Kaparulin}, and classical rigid string \cite{Polyakov,Curtright,Simon}. Interestingly, the model is also applied to polymer physics \cite{Harris}, formation of microemulsions \cite{DeGennes}, and membrane biology \cite{Canham,Helfrich,Kleinert2}.
However, to my knowledge, there is no article dealing with relativistic higher derivative Lagrangians of this form:
\begin{equation}\label{L}
\widetilde{L}(\dot{x},\ddot{x},\cdots,x^{(n)})\\
=\alpha_1 A_{\mu}(x)\dot{x}^\mu+\alpha_2 G_{\mu}(x)\ddot{x}^\mu+\cdots+\alpha_n U_{(n)\mu}(x)x^{(n)\mu} 
\end{equation}
where $x^{(n)}(s)\equiv d^n x(s)/ds^n$ are the $n$-derivatives of the position ($ds=cd\tau$, where $\tau$ is the proper time),
and $U_{(n)\mu}(x),\ n=1,2,3,..$ are the generalized vector -fields coupling linearly with the $x^{(n)}$ vectors.
Here, we denote the field $U_{(1)\mu}(x)=A_\mu(x)$ to refer to the electromagnetic potential and $U_{(2)\mu}(x)=G_\mu(x)$ by analogy with the geodesic equations we obtain in equation (\ref{EqGeodesique}). In this article, we first derive the equations of motion of the a massive particle experiencing interactions described by the Lagrangians of the form \eqref{L} and show that odd and even orders Lagrangians are analogous to the electromagnetic and the gravitational interaction, respectively.   
Next, we generalize the concept of electromagnetic induction assuming the coupling between the higher fields $U_{(n),\mu}(x),\ n\geq1$
with the \textit{higher derivative currents} $j^{(n)\mu}=\rho(x)d^nx^{\mu}/ds^n$, where $\rho(x)$ is the spatial density of mass ($n$ even) or of electric charge ($n$ odd). In the last section, we discuss the applications to the model to microscopic physics and general relativity.

\section{Equations of motion of a massive particle}

In this section, we analyze the equations of motion of a massive particle experiencing the  
action defined by
\begin{equation*}
S=\int ds L_0(\dot{x})+\int ds  \widetilde{L}(\dot{x},\ddot{x},\cdots,x^{(n)}) \ ,
\end{equation*}
where $L_0(\dot{x})\equiv\frac{mc^2}{2}\dot{x}_\mu\dot{x}^\mu$ and where the interaction Lagrangian $\widetilde{L}$ is given by equation \eqref{L}.

\subsection{Second and third order Lagrangians}

First, we consider the case $n=2$ and $n=3$. 
By integration by parts for $n=2$ we find the following action \cite{Anderson1}
\begin{equation}\label{Srew}
\widetilde{S}=\alpha_1\int ds A_{\mu}(x)\dot{x}^\mu - \alpha_2 \int ds\ \left(\partial_\nu G_{\mu}\right)\dot{x}^\mu\dot{x}^\nu\ ,
\end{equation}
and one can see that the first part of the action $\widetilde{S}$ is similar to the electrodynamics action
whereas the second part is similar to the gravitational action.
Indeed, from the generalized Euler-Lagrange equations 
(see \cite{Borneas1,Borneas2,Anderson1,Anderson2,Alder}):
\begin{eqnarray}\label{Euler-Lagrange}
\frac{d^2}{ds^2}\left(\frac{\partial L}{\partial \ddot{x}^\mu}\right)-
\frac{d}{ds}\left(\frac{\partial L}{\partial \dot{x}^\mu}\right)+\frac{\partial L}{\partial x^\mu}=0\ ,
\end{eqnarray}
with $L=L_0+\widetilde{L}$, we obtain
\begin{eqnarray}\label{EqGeodesique}
mc^2\eta_{\mu\nu}\ddot{x}^\nu-\alpha_2\varepsilon_{\mu\nu}\ddot{x}^\nu-\alpha_2\Delta_{\mu\nu\sigma}\dot{x}^\nu\dot{x}^\sigma
=-\alpha_1 F_{\mu\nu}\dot{x}^\nu\ ,
\end{eqnarray}
where $\varepsilon_{\mu\nu}$ and $\Delta_{\mu\nu\sigma}$ are defined by
\begin{subequations}\label{Eq:n2:def}
\begin{eqnarray}
&\varepsilon_{\mu\nu}\equiv\partial_\mu G_\nu + \partial_\nu G_\mu \label{g}\ ,\\
&\Delta_{\mu\nu\sigma}\equiv\partial_\nu\partial_\sigma G_\mu 
=\frac{1}{2}(\partial_\nu \varepsilon_{\mu\sigma}+
\partial_\sigma \varepsilon_{\mu\nu}-\partial_\mu \varepsilon_{\nu\sigma}) \label{gamma1}\ ,
\end{eqnarray}
\end{subequations}
and where $F_{\mu\nu} \equiv \partial_\mu A_\nu - \partial_\nu A_\mu$ has a mathematical form that is similar to the Faraday tensor.
From equations \eqref{EqGeodesique} and \eqref{Eq:n2:def}, one can see the analogy with the equations of the motion of a charged particle in a gravitational field and in an electromagnetic field.
However, the fixed metric (or \textit{background} metric) is Minkowskian. 
Then, the quadrivector field $G_\mu(x)$ can be seen as a \textit{displacement vector field} 
and $\varepsilon_{\mu\nu}$ can be viewed as an \textit{strain tensor}
by analogy with the deformation theory of a continuous medium \cite{MMC}, \cite{Beau}.

Now, we take $n=3$ and we denote $U_{(3)\mu}\equiv B_\mu$. From the generalized Euler-Lagrange equations
\begin{eqnarray}\label{Euler-Lagrange3}
-\frac{d^3}{ds^3}\left(\frac{\partial L}{\partial \dddot{x}^\mu}\right)+\frac{d^2}{ds^2}\left(\frac{\partial L}{\partial \ddot{x}^\mu}\right)-
\frac{d}{ds}\left(\frac{\partial L}{\partial \dot{x}^\mu}\right)+\frac{\partial L}{\partial x^\mu}=0\ ,
\end{eqnarray}
we derive the following equations of motion
\begin{equation}\label{EqGeodesique3}
mc^2\eta_{\mu\nu}\ddot{x}^\nu - \alpha_2\varepsilon_{\mu\nu}\ddot{x}^\nu  -\alpha_2\Delta_{\mu\nu\sigma}\dot{x}^\nu\dot{x}^\sigma  +\alpha_3 H_{\mu\nu}\dddot{x}^\nu-\alpha_3\Upsilon_{\mu\nu\sigma\rho}\dot{x}^\nu\dot{x}^\sigma\dot{x}^\rho
-3\alpha_3\varSigma_{\mu\nu\sigma}\ddot{x}^\nu\dot{x}^\sigma
=-\alpha_1 F_{\mu\nu}\dot{x}^\nu \ ,
\end{equation}
where 
\begin{equation}
\begin{cases}
H_{\mu\nu}\equiv\partial_\mu B_\nu - \partial_\nu B_\mu \label{g}  \\
\varSigma_{\mu\nu\sigma}\equiv\partial_\nu\partial_\sigma B_\mu \\
\Upsilon_{\mu\nu\sigma\rho}\equiv\partial_\nu\partial_\sigma\partial_\rho B_\mu
\end{cases} \ .  
\end{equation}
We can see that this field generalizes the idea of the electromagnetic field because of $H_{\mu\nu}$ is antisymmetric.
However, in (\ref{EqGeodesique3}) there are some other fields, similar to $\Delta_{\mu\nu\sigma}$, coupling with the combinations
of the odd derivatives of $x^{\mu}$, i.e. $\ddot{x}^\nu\dot{x}^\sigma$ and $\dot{x}^\nu\dot{x}^\sigma\dot{x}^\rho$.

\subsection{Higher order Lagrangians}

Now, we shall discuss the higher derivative terms.
For $n=4$, we denote the field $K_\mu(x)\equiv U_{(4)\mu}(x)$. 
The dynamic equations have a similar structure to the one we obtained for $G_\mu(x)$ (i.e. for $n=2$).
As it has been shown that in the non-relativistic theory \cite{Anderson1}, we notice that the Lagrangian $\alpha_4 x_\mu\ddddot{x}^\mu$ 
is equivalent to this Lagrangian $\alpha_4 \ddot{x}_\mu\ddot{x}^\mu$ 
and the quantity $\alpha_2\dot{x}^2+\alpha_4\ddot{x}_\mu\ddot{x}^\mu-2\alpha_4\dot{x}_\mu\dddot{x}^\mu$ 
could be interpreted as a more \textit{general kinetic energy} \cite{Alder}.
Similarly, the Lagrangian $\alpha_4 K_{\mu}(x)\ddddot{x}^\mu$ 
is equivalent to $\alpha_4\left(\partial_{\nu}K_\mu\right) \ddot{x}^\mu\ddot{x}^\nu
+\alpha_4\left(\partial_{\sigma}\partial_{\nu}K_\mu\right) \ddot{x}^\mu\dot{x}^\nu\dot{x}^\sigma$, which
has a more complicated expression than the one we obtained for the special case $K_\mu(x)=x_\mu$.

To finish this section, let us now consider the generalized fields $U_{(n)\mu}(x),\ n\geq1$. 
From the generalized Euler-Lagrange equations
$$
\sum_{k=0}^{n}(-1)^k \frac{d^k}{ds^k}\frac{\partial L}{\partial x^{(k)\mu}}=0\ ,
$$
we find terms of the form $\partial_{\mu_{1}}\cdots \partial_{\mu_{p}}U_{(n)},\ p=1,\cdots,n$ 
multiplied by the combination of the derivatives $x^{(l_1)\mu_1}x^{(l_2)\mu_2}\cdots x^{(l_p)\mu_p}$, $\sum_{j=1}^{p}l_j=n$.   
The equations of motion shows that the ``even'' $n$-fields are analogous to the "gravitational field" 
while the ``odd'' $n$-fields are analogous to the "electromagnetic field"
as the derivatives $x^{(n)\mu}$ are multiplied by the 
symmetric (if $n$ is even) / antisymmetric (if $n$ is odd) part of the 
first derivative of the field:
$$(\partial_{\mu} U_{\nu}+(-1)^n\partial_{\nu}U_{\mu})x^{(n)\nu}\ .$$
We will see the consequences of this remark in the next section.

\section{General fields hypothesis}

In this section, we propose to formulate a dynamical theory of the generalized vector fields introduced above. We introduce a \textit{generalized induction principle}, analogous to the electromagnetic induction.

\subsection{Construction of the n=2-field equations by analogy with the vector electromagnetic field}

In equation (\ref{L}) for $n=2$ we notice that the field $G_\mu$ is coupled with the acceleration of the particle
as the field $A_{\mu}$ is coupled with the velocity of the particle.
By analogy with the construction of the electromagnetic field theory, we suggest the following field equations:  
\begin{equation}\label{gpe2}
\partial_\mu \varepsilon^{\mu\nu}(x) = -\kappa j^{(2)\nu}(x)\ ,
\end{equation}
where the acceleration current density $j^{(2)\nu}$ (generally non-conserved) is: 
\begin{equation}\label{QuadridensitAccRG}
j^{(2)\nu}(x)\equiv \rho_m(x)c^2\frac{du^\nu}{ds}\ ,
\end{equation}
where $\rho_m(x)$ is the density of particles and $\frac{du^\nu}{ds}$ is the 4-acceleration. This equation is the analog of the Gauss-Ampère law in electrodynamics $\partial_\mu F^{\mu\nu} = \mu_0 j^{(1)\nu}$, where $j^{(1)\nu}$ is the 4-electric current density. 
Notice that we can rewrite the coupling constant $\kappa$ as follows 
$$\kappa=\frac{8\pi G \lambda^2}{c^4}$$ 
where $\lambda$ has the dimension of a length.  

To complete the system of field equations, we need ten additional equations: 
\begin{equation}\label{gpe1}
\partial_\sigma \partial^\sigma \varepsilon_{\mu\nu}+\partial_\mu \partial_\nu \varepsilon_{\sigma}^{\sigma}
=\partial_\mu \partial^\sigma \varepsilon_{\sigma\nu}+\partial_\nu \partial^\sigma \varepsilon_{\sigma\nu}\ ,
\end{equation}
The equations (\ref{gpe1}) have the same form as the compatibility equations 
for the strain tensor in the three-dimensional non-relativistic theory of deformation of continuous media \cite{MMC,Beau}. 
There are also the analog of the Gauss-Faraday equations $\partial_\mu F_{\nu\sigma}+\partial_\nu F_{\sigma\mu}+\partial_\sigma F_{\mu\nu}$.

After combining the last two groups of equations \eqref{gpe2}-\eqref{gpe1}, we can easily derive the wave equations
\begin{eqnarray}\label{PropaChampsPhys0}
\Box \varepsilon_{\nu\sigma}(x)+\partial_{\nu}\partial_\sigma \varepsilon_\mu^\mu(x)=-\kappa \xi_{\nu\sigma}^{(2)}(x) \ ,
\end{eqnarray}
where $\xi^{(2)}_{\nu\sigma}(x)\equiv\partial_\sigma j^{(2)}_\nu(x)+\partial_\nu j^{(2)}_\sigma(x)$ .
Also, the trace of $\varepsilon_{\mu\nu}$ satisfies the equation 
\begin{eqnarray}\label{Tracen2}
\Box \varepsilon_{\mu}^\mu(x)=-\kappa\partial_\mu j^{(2)\mu}(x)\ , 
\end{eqnarray}
this means that $\varepsilon_\mu^\mu$ is a non-massive scalar field.
It is worth mentioning that the trace of the strain tensor is usually interpreted as the contraction/dilation of the volume of the continuous medium \cite{MMC}.
Hence, from equation (\ref{Tracen2})
we conclude that the relativistic deformation of the volume of a \textit{four dimensional continuous medium} 
is related to the non-conservation of the current $j^{(2)}$.

\subsection{Generalization to the n-field equations}

Now, we can generalize the construction of the general field theory for any $n\geq1$.
Following the rules mentioned above, we rewrite the constants in (\ref{L}) as $\alpha_{2n}=mc^2(\lambda_{n})^{2n}$ 
and $\alpha_{2n-1}=q\frac{(\xi_{n})^{2n-2}}{c^{2n-2}},\ n\geq1$,
where $\lambda_n$ and $\xi_n$ are fundamental constants that have the dimension of length and $G$ is the universal gravitational constant. We denote $m$ and $q$ the mass and the electric charge of the particle, respectively. 

It comes naturally that for so-called \textit{gravitational-type fields} $U_{(2n)}\equiv G_{(n)},\ n\geq1$,
the coupling has the form:
\begin{equation}
-\frac{8\pi G}{c^2}\frac{(\lambda_{n})^{2n}}{c^{2n}}G_{(n)\mu}(x)j^{(2n)\mu}(x)\ ,
\end{equation}
whereas for the so-called \textit{electromagnetic-type fields} $U_{(2n-1)}\equiv A_{(n)},\ n\geq1$, the coupling reads:
\begin{equation}
\mu_0\frac{(\xi_{n})^{2n-2}}{c^{2n-2}}A_{(n)\mu}(x)j^{(2n-1)\mu}(x)\ ,
\end{equation}
where $A_{(n)\mu}$ has the dimension of $\mathrm{N.A^{-1}}$ ($\mathrm{N}$ is the Newton and $\mathrm{A}$ the Amp\`ere),
$\mu_0$ is the vacuum permeability ($\mu_0=4\pi\times 10^{-7}\ \mathrm{N.A^{-2}}$),
and where $G_{(n)\mu}$ has the dimension of a length.
The generalized currents for $n=1,2,3,..$ are constructed as follows:

\begin{subequations}\label{GenCurrent}
\begin{equation}\label{GenCurrenteven}
j^{(n)\nu}\equiv  \rho_m(x)\frac{d^{n}x^\nu}{d\tau^{n}},\ \mathrm{if\ n\ is\ even} \ ,
\end{equation}
\begin{equation}\label{GenCurrentodd}
j^{(n)\nu}\equiv   \rho_e(x)\frac{d^{n}x^\nu}{d\tau^{n}},\ \mathrm{if\ n\ is\ odd} \ ,
\end{equation}
\end{subequations}

where $\rho_m(x)$ is the mass density and $\rho_e(x)$ the electric charge density.
Similarly to equation (\ref{gpe2}), we construct an $(2n-1)$-order linear differential theory to relate the sources and the fields:
\begin{subequations}
\begin{equation}\label{gpe2n}
O^{(n)}_\mu(\lambda_n) \epsilon_{(n)}^{\mu\nu}(x) = - \frac{8\pi G}{c^2}\frac{\lambda_n^{2n}}{c^{2n}} j^{(2n)\nu}(x),\ n\geq1\ ,
\end{equation}
\begin{equation}\label{gpe2n}
Q^{(n)}_\mu(\xi_n) f_{(n)}^{\mu\nu}(x) = - \mu_0\frac{\xi_n^{2n-2}}{c^{2n-2}} j^{(2n-1)\nu}(x),\ n\geq1\ ,
\end{equation}
\end{subequations}
where $O^{(n)}_\mu(\lambda_n)$ and $Q^{(n)}_\mu(\xi_n)$ are two $(2n-1)$-order differential operators 
and where $\varepsilon_{(n)}^{\mu\nu}(x)\equiv \partial^\mu G_{(n)}^\nu+\partial^\nu G_{(n)}^\mu$
and $f_{(n)}^{\mu\nu}(x)\equiv \partial^\mu A_{(n)}^\nu-\partial^\nu A_{(n)}^\mu$. 

From those rules we can obtain similar wave equations to \eqref{PropaChampsPhys0} and \eqref{Tracen2} with higher order differential operators $(\lambda_n)^{2k}\underbrace{\Box\Box\cdots\Box}_{k\ \mathrm{times}},\ k=1,\cdots,n$. 
For example, for $n=4$, we can take $O^{(4)}_\mu(\lambda)=(\lambda^2\Box+1)\partial_\mu$ and
then we get the wave equation for the trace of the tensor $\zeta_{\mu\nu}\equiv \partial_\mu K_\nu+\partial_\nu K_\mu$: 
\begin{eqnarray}\label{Tracen4}
\left(\lambda^2\Box+1\right)\Box \zeta_{\mu}^\mu(x)=- \frac{8\pi G\lambda^4}{c^6} \partial_\mu j^{(4)\mu}(x)\ ,
\end{eqnarray} 
and then $\Box\zeta_\mu^\mu$ is a massive scalar field. Similar equations have been studied in the context of generalized electrodynamics \cite{PodolskyI} and of higher derivative scalar field theories \cite{Anisimov,Kaparulin}. The difference here is that the source of the scalar field is related to the fourth-order general current $j^{(4)\mu}(x)$, which is proportional to the fourth order derivative of the position (i.e., to the so-called \textit{snap}), see equation \eqref{GenCurrent}. 

Notice that for the electromagnetic-type fields, the choice of the field $A_{(n)\mu}$ is not unique because the fields $f_{(n)}^{\mu\nu}$ is antisymmetric. On the contrary, all of the components of the gravitational-like field $G_{(n)\mu}$ are physical because the tensor $\varepsilon^{(n)}_{\mu\nu}$ is symmetrical.

\subsection{Unitary fields}\label{Unitary} 

Physically, the \textit{generalized vector fields} can be understood 
as a $n-1$th order perturbative correction of the electromagnetic and gravitational vector theory,    
i.e. $A_{(2)\mu}\equiv B_\mu(x)$ is the first order correction of the Minkowskian theory of Electromagnetism field $A_{(1)\mu}\equiv A_\mu$.
Hence, it becomes natural to unify the gravity type fields as well as the electromagnetic type fields.
In order to unify the fields, we construct the dimensionless unification constants: 
$$\gamma_{jl}=\frac{\lambda_j}{\lambda_l},\ \theta_{jl}=\frac{\xi_j}{\xi_l},\ j,l=1,2,3,\cdots$$
where the constants $\lambda_n,\ n\geq1$ and $\xi_{n},\ n\geq1$,
were introduced in the previous section.

For example, if we suppose that $A_\mu(x)=B_\mu(x)$, we get the coupling:
$$\mu_0 A_{\mu}(x)\left( j^{(1)\mu}(x)+\frac{\xi^2}{c^2}j^{(3)\mu}(x)\right)$$
where we put $\xi_2=\xi$ (we recall that $\alpha_1=1$ and $\alpha_3=\xi_2^2/c^2$).
Phenomenologically, this means that for an electric circuit with an intensity of this type $I(t)=I_0 t^2/\tau^2$,
where $\tau$ is a time constant, the third order time derivative of the vector position of the electrons in the current is non-zero 
(this kinematic quantity is called the \textit{Jerk}, see \cite{Jerk1,Jerk2})
and so that the electromagnetic field would be modified by the \textit{jerk current} $j^{(3)}$. 
We mention that in the generalized theory of Electrodynamics \cite{PodolskyI}
the relation with the higher derivatives currents (c.f. eqs. \eqref{GenCurrentodd}, \eqref{gpe2n}) has not been suggested. 

Similarly, we can construct the unified coupling for the even fields $n=2$ and $n=4$
$$-\frac{8\pi G}{c^2}\frac{\lambda^{2}}{c^{2}}G_{\mu}(x)\left( j^{(2)\mu}(x)+\gamma^4\frac{\lambda^2}{c^2}j^{(4)\mu}(x)\right)$$
assuming that $G_\mu(x) = K_\mu(x) \equiv G_{(2)}(x)$, where we put $\lambda_1=\lambda,\ \gamma=\gamma_{21}$, and where we introduce the effective current
$\widetilde{j}^{(2)\mu}=j^{(2)\mu}+\gamma^4\frac{\lambda^2}{c^2}j^{(4)\mu}$.
Consequently, the effective strain tensor $\epsilon_{\mu\nu}$ that deforms the Minkowski metric
is also induced by the second order derivative of the acceleration (the so-called \textit{snap}) of the moving particles.

\section{Discussion}

\subsection{Microscopic Physics and generalized currents}

The effect of gravitation at the microscopic scale is not yet well known. It is also clear that the geometrical description of gravity fails at the Planck scale \cite{Simon}. Therefore, it is fair to ask whether the current density of accelerated masses $j^{(2)}$ plays a significant role at this scale. 

Theoretically, it could be interesting to investigate the effect of the generalized currents $j^{(n)}$ on the electromagnetic and gravitational fields. We can assume that the value of the parameters $\xi$ and $\lambda$ introduced in Section \ref{Unitary} are very small, for the physical effects of generalized currents have never been observed. However, higher order derivatives should play a role at the Planck scale \cite{Simon,Simon2}. Hence, our theory of generalized fields and forces could be relevant in high-energy particle physics.
Formulating a generalized quantum field theory is a challenging problem for future investigations.

\subsection{Strain and stress tensor in General Relativity}

The special case $n=2$ (we consider $\alpha_{n\geq3}=0$) has been recently discussed in \cite{Beau,Beau2}. It consists of a modification of the Einstein general relativity theory (GR) that deserves to be mentioned in this paper.
In this section we introduce the idea of a covariant strain/stress field theory in the framework of GR.

Similarly to \eqref{Eq:n2:def}, it is clear that the strain tensor in Riemann spaces is given by $\varepsilon_{\mu\nu}=D_\mu G_\nu+D_\nu G_\mu$ where the operator $D$ is the covariant derivative. We construct a stress tensor
\begin{equation}\label{Sigma}
\sigma_{\mu\nu}(x)=\alpha g_{\mu\nu}\varepsilon_{\gamma}^{\gamma}+2\beta \varepsilon_{\mu\nu}
\end{equation}
where $\alpha$ and $\beta$ are the Lam\'{e} coefficient, and where the trace $\varepsilon_{\gamma}^{\gamma}$ corresponds to the relative variation of the four-dimensional volume $\delta V/V$ of the space-time continuum. If one considers a hydrostatic fluid without shear stress (i.e., $\beta=0$), the modulus $\alpha$ corresponds to the bulk modulus of the medium. 

In this framework, the cosmological term $\Lambda g_{\mu\nu}$ in the standard model of cosmology can be interpreted as a deformation of a 4-dimensional medium, where $B=-\rho_G c^2=\frac{\Lambda c^4}{8\pi G}$ corresponds to the bulk modulus for an isotropic medium.
The stress tensor \eqref{Sigma} is then added to the Einstein field equations of gravity:
\begin{equation}\label{GR}
R_{\mu\nu}-Rg_{\mu\nu}=\frac{8\pi G}{c^4}\left(T_{\mu\nu}+\sigma_{\mu\nu}\right)
\end{equation}
and subsequently we obtain the relations:
\begin{equation}\label{Conserv}
D_{\mu}\sigma^{\mu\nu}(x)+D_\mu T^{\mu\nu}(x)=0 
\end{equation}
where $T_{\mu\nu}$ is the energy-momentum tensor in the Einstein field equation. 
This equation means that the total energy in the universe is conserved but that the ordinary matter-energy can be accelerated. Notice that for a general stress tensor $\sigma_{\mu\nu}\equiv C_{\mu\nu\gamma\delta}\varepsilon^{\gamma\delta}$, equation \eqref{Conserv} generalizes \eqref{gpe2} (where for the sake of simplicity we put $\alpha=0$ and $2\beta=\kappa^{-1}$).
Gathering \eqref{GR} and \eqref{Conserv} with the equation of fields in $T_{\mu\nu}$, we obtain an incomplete set of equations. Thus, there must exist an additional coupling between the matter-energy $T_{\mu\nu}$ and the strain field $\varepsilon_{\mu\nu}$. The construction of this coupling has been discussed for Minkowski space in the previous sections and generalized for GR in \cite{Beau} and \cite{Beau2}. The idea is the following, the effective strain tensor $\varepsilon_{\mu\nu}$ deforms (linearly) the Lagrangian in such a way that it introduces a coupling $\varepsilon_{\mu\nu}T^{\mu\nu}$ as in \eqref{Srew}. Hence, the effect of this additional term is to modify the standard field equations by modifying the covariant derivative $D$ by a linear deformation similar to the one obtained in equations \eqref{EqGeodesique}, \eqref{EqGeodesique3} (deformations of $\eta$ by $\varepsilon$).   
To conclude the variation of inertia of the ordinary matter-energy \eqref{Conserv} is compensated by the divergence of the stress energy of the continuous medium. Somehow, this extension of GR revisits Mach's principle as the inertia (of the matter-energy) also depends on the deformation of the curved space-time that is induced by the matter-energy contained therein.
We refer the reader to the paper of Einstein on a related topic \cite{Einstein}. 
The relativistic theory of an Aether was discussed several times, see for e.g. \cite{Aether1}, \cite{Aether2}.
In the present article, our hypothesis is different and gives a relativistic theory of the deformation of continuous media, for which the geometry is still described by the metric field whereas the strain tensor is an additional field.

Beyond $n=2$, the strain/stress interpretation is no longer valid and requires an other kind of generalization of the theory of relativity. This generalization should revisit the concept of inertia by including higher order derivatives of the vector position, as shown in Section 2. It should also include sources with higher derivatives, as shown in Section 3. For instance, the energy-momentum tensor of a distribution of massive particle at the order of $\lambda^2$ should read 
$$
\frac{1}{2}\rho\lambda^2   \frac{D{u}^\mu}{Ds}\frac{D{u}^\nu}{Ds}\ ,
$$
where $\rho$ is the density of mass, $u^\mu$ is the covariant velocity and $D{u}^\mu/Ds$ is the covariant acceleration (this is the coviariant version of the acceleration energy term $(m\lambda^2/2)\ddot{x}^2$ in the Pais-Uhlenbeck oscillator, \cite{Pais,Bolonek}). This is beyond the scope of this article, however, we hope this novel way of generalizing relativity will inspire future research.

\end{document}